\documentclass[12pt,english]{rspublic}
\usepackage[T1]{fontenc}
\usepackage{latexsym, amssymb, enumerate, amsmath}
\usepackage{epsfig}
\usepackage{color}
\usepackage{hyperref}
\usepackage{changebar}
\usepackage{babel}
\usepackage{tikz}
\usetikzlibrary{decorations.pathreplacing}
\usepackage{natbib}
\usetikzlibrary{decorations}
\usetikzlibrary{decorations.pathmorphing}

    \usepackage{graphicx}
= 0.5\thickmuskip \arraycolsep = 0.3\arraycolsep

\definecolor{dgreen}{rgb}{0,.5,0}
\definecolor{grau}{gray}{.5}
\definecolor{schwarz}{gray}{0}

\newcommand{\reff}[1]{(\ref{#1})}
\newcommand{\ol}[1]{\overline{#1}}

\newcommand{\cg}[1]{\mathcal{#1}}

\newcommand{\av}[1]{\left|#1\right|}

\newcommand{\brkts}[1]{\left(#1\right)}
\newcommand{\ebrkts}[1]{\left[#1\right]}

\newcommand{\brcs}[1]{\left\{#1\right\}}

\newcommand{\pd}[2]{\frac{\partial #1}{\partial #2}}

\newcommand{\bsplitl}[2]{
\begin{equation}
\begin{split}
#1
\end{split}
\label{#2}
\end{equation}}

\newtheorem{thm}{Theorem}[section]

\newtheorem{cor}[thm]{Corollary}

\newtheorem{defn}[thm]{Definition}

\newtheorem{rem}[thm]{Remark}

\DeclareGraphicsExtensions{.png,.eps,.ps}
\graphicspath{{archive_pics/}}

\begin{document}

\title[Upscaled phase-field models]{Upscaled phase-field models for interfacial dynamics in strongly heterogeneous domains}
\author[M. Schmuck, M. Pradas, G. A.  Pavliotis, and S. Kalliadasis]{Markus Schmuck$^{1,2,\thanks{Author for correspondence (m.schmuck@imperial.ac.uk).}}$, Marc Pradas$^1$, Greg A. Pavliotis$^{2,3}$, and Serafim Kalliadasis$^{1}$}
\affiliation{$^1$ Department of Chemical Engineering, Imperial College London,
South Kensington Campus, SW7 2AZ London, UK\\
$^2$ Department of Mathematics, Imperial College London,
South Kensington Campus, SW7 2AZ London, UK\\
$^3$ CERMICS, Ecole Nationale des Ponts et Chauss\'ees,
   6 \& 8 Avenue Blaise Pascal, 77455 Marne La Vall\'ee Cedex 2, FR}

\maketitle

\begin{abstract}{Phase-field models, Cahn-Hilliard equation, multiscale modeling, homogenization, porous media, wetting}
 We derive a new effective macroscopic Cahn-Hilliard equation whose homogeneous free energy is 
represented by 4-th order polynomials, 
which form the frequently applied double-well potential. 
This upscaling is done for perforated/strongly heterogeneous domains. 
To the best knowledge of the authors, this seems to be the
first attempt of upscaling the Cahn-Hilliard equation in such
domains. The new homogenized equation should have a broad range of
applicability due to the well-known versatility of phase-field
models. The additionally introduced feature of systematically and
reliably accounting for confined geometries by homogenization allows
for new modeling and numerical perspectives in both, science and
engineering. Our results are applied to wetting dynamics in porous
media and to a single channel with strongly heterogeneous walls.
\end{abstract}

\section{Introduction}\label{sec:Intr}
Consider the abstract energy density
\bsplitl{
e(\phi)
    := F(\phi) 
    +\frac{\lambda^2}{2}\av{\nabla\phi}^2\,,
}{FrEn} where $\phi$ is a conserved density that plays the role of
an order-parameter by taking appropriate equilibrium limiting values
that represent different phases. The gradient term
$\lambda^2\av{\nabla\phi}^2$ penalizes the interfacial area between
these phases, and the bulk free energy $F$ is defined as the
polynomial \bsplitl{ F(\phi)
    & := \int_0^\phi f(s)\,ds\,,
\quad\textrm{and}\quad f(s):= a_3s^3+a_2s^2+a_1s\,. }{hFrEn} In the
Ginzburg-Landau/Cahn-Hilliard formulation, the total energy is defined by
$E(\phi):=\int_\Omega e(\phi)\,d{\bf x}$ with density \reff{FrEn}
on a bounded $C^{1,1}$-domain $\Omega\subset\mathbb{R}^d$ with $1\leq d\leq 3$
denoting the spatial dimension. In general, the local
minima of $F$ correspond to the equilibrium limiting values of
$\phi$ representing different phases separated by a diffuse
interface whose spatial extension is governed by the gradient term.

It is well accepted that thermodynamic equilibrium can be achieved
by minimizing the free energy $E$, here supplemented by a possible boundary
contribution $\int_{\partial\Omega}g({\bf x})\,do({\bf x})$ for $g({\bf x})\in H^{3/2}(\partial\Omega)$,
with respect to its gradient flow over the domain $\Omega$, that means,
\bsplitl{
\textrm{(Homogeneous case)}\,\,\,
\begin{cases}
\pd{}{t}\phi
    = {\rm div}\brkts{
    \hat{\rm M}\nabla\brkts{
        f(\phi) 
        -\lambda^2\Delta\phi
        }
    }
    & \quad\textrm{in }\Omega_T\,,
\\
\nabla_n\phi:= {\bf n}\cdot\nabla\phi
    = g({\bf x})
    & \quad\textrm{on }\partial\Omega_T\,,
    \,,
\\
\nabla_n\Delta\phi
    = 0
    & \quad\textrm{on }\partial\Omega_T
    \,,
\end{cases}
}{PhMo}
where $\Omega_T:=\Omega\times]0,T[$, $\partial\Omega_T:=\partial\Omega\times]0,T[$, $\phi$ satisfies the initial condition $\phi({\bf x},0)
    = \psi({\bf x})$, and $\hat{\rm M}=\brcs{{\rm m}_{ij}}_{1\leq i,j\leq d}$ denotes a mobility tensor with real and bounded elements ${\rm m}_{ij}>0$. Equation \reff{PhMo} is the
gradient flow with respect to the $H^{-1}$-norm, here weighted by
the mobility tensor $\hat{\rm M}$, and is referred to as the
Cahn-Hilliard equation. This equation is a model prototype for interfacial dynamics [e.g.
\cite{Fife1991}] and phase transformation [e.g. \cite{Cahn1958}] under homogeneous Neumann boundary conditions,
i.e., $g=0$, and a free energy $F$ representing the phenomenological standard
double-well potential $F(s)=\frac{1}{4}\brkts{s^2-1}^2$. The
polynomial $f=F'$, defined in \reff{hFrEn}, encloses a set of free
energies which allow for the same steps in the rigorous
homogenization process leading to the main result of this paper, Theorem \ref{thm:EfPhFi}.
We emphasize that $F$ represents a bulk free energy which is
well-accepted since it allows for stable numerics and captures
phenomenoligically the features of systematically derived free
energies such as the regular solution model  [e.g. \cite{Cahn1958}] based on the free energy
of mixing, i.e., 
\bsplitl{
f(\phi)=kT\brkts{
        \phi{\rm ln}\phi
        +(1-\phi){\rm ln}(1-\phi)
    }
    +a\phi(1-\phi)\,.
}{ReSoMo} The mean free energy \reff{ReSoMo} can be derived by a
thermodynamic limit from lattice gas models of filled and empty
sites for instance. Unfortunately, the energy \reff{FrEn} cannot be
reduced to the atomistic Lennard-Jones potential. But \reff{FrEn} is
related to the Lennard-Jones potential in the sense of the LMP
(Lebowitz, Mazel and Presutti) theory \cite[]{Presutti2009}.
It is well-known, that formally, the energy \reff{FrEn} dissipates along solutions of the gradient flow~\reff{PhMo}, that means, 
$E(\phi(\cdot,t))
    \leq
    E(\phi(\cdot,0))
    =:E_0\,.
$ 
This follows immediately after differentiating \reff{FrEn} with respect to time and using
\reff{PhMo} for $g=0$.

\medskip


\begin{figure}[htbp]
\begin{center}


\setlength{\unitlength}{1cm}
\begin{picture}(11.75,7.3)(0,0)                    

\setlength{\unitlength}{0.1cm}
\setlength{\unitlength}{0.9cm}
\thicklines
\multiput(0.0,0.0)(0.8,0.0){6}{\framebox(0.8,0.8)[s]{\put(0.4,0.0){\circle*{0.6}}}}
\multiput(0.0,0.0)(0.8,0.0){6}{\framebox(0.8,0.8)[s]{\put(0.2,0.0){\textcolor{white}{$Y^2$}}}}
\multiput(0.0,0.8)(0.8,0.0){6}{\framebox(0.8,0.8)[s]{\put(0.4,0){\circle*{0.6}}}}
\multiput(0.0,0.8)(0.8,0.0){6}{\framebox(0.8,0.8)[s]{\put(0.2,0){\textcolor{white}{$Y^2$}}}}
\put(4.8,-0.015){\line(1,0){0.5}}
\multiput(0.0,1.6)(0.8,0.0){6}{\framebox(0.8,0.8)[s]{\put(0.4,0){\circle*{0.6}}}}
\multiput(0.0,1.6)(0.8,0.0){6}{\framebox(0.8,0.8)[s]{\put(0.2,0){\textcolor{white}{$Y^2$}}}}
\put(4.8,0.8){\line(1,0){0.5}}
\put(5.05,0.0){\vector(0,1){0.8}}
\put(5.05,0.8){\vector(0,-1){0.8}}
\put(5.15,0.35){$\ell$}
\multiput(0.0,2.4)(0.8,0.0){6}{\framebox(0.8,0.8)[s]{\put(0.4,0){\circle*{0.6}}}}
\multiput(0.0,2.4)(0.8,0.0){6}{\framebox(0.8,0.8)[s]{\put(0.2,0){\textcolor{white}{$Y^2$}}}}
\multiput(0.0,3.2)(0.8,0.0){6}{\framebox(0.8,0.8)[s]{\put(0.4,0){\circle*{0.6}}}}
\multiput(0.0,3.2)(0.8,0.0){6}{\framebox(0.8,0.8)[s]{\put(0.2,0){\textcolor{white}{$Y^2$}}}}
\put(0.0,4.3){Periodic covering by cells $Y$}
\put(5.6,2.25){($\epsilon\to 0$)}
\put(5.1,2.0){\vector(2,0){2.0}}
\put(7.5,0.0){\fcolorbox{black}{grau}{\makebox(4.8,3.9)[]{}}} %
\put(7.75, 4.3){Homogenous approximation}
\put(9.8, 1.9){$\Omega$}

\put(12.53,4.005){\line(1,0){0.6}}
\put(12.53,-0.105){\line(1,0){0.6}}
\put(12.8,2){\vector(0,1){2.0}}
\put(12.8,2.0){\vector(0,-1){2.1}}
\put(13.0,1.9){$L$}

\put(2.2,6.1){Reference cell $Y$}

\put(7.15,7.45){\line(1,0){0.7}}
\put(7.15,5.215){\line(1,0){0.7}}
\put(7.5,6.2){\vector(0,1){1.25}}
\put(7.5,6.25){\vector(0,-1){1.02}}
\put(7.65,6.2){$\ell$}
\setlength{\unitlength}{0.9cm}
\put(5,5.2){
\begin{tikzpicture}
\draw[fill=black] (15,15) circle (.8cm); \draw[line width=0.6mm]
(14,14) rectangle (16,16);
\end{tikzpicture}
}
\put(6.1,6.5){\textcolor{white}{$Y^2$}}
\put(5.55,5.9){\textcolor{white}{$:=Y\setminus Y^1$}}

\end{picture}
\caption{{\bf Left:} Strongly heterogeneous/perforated material as a periodic covering of reference cells $Y:=[0,\ell]^d$.
{\bf Top, middle:} Definition of the reference cell $Y=Y^1\cup Y^2$ with $\ell=1$.
{\bf Right:} The ``homogenization limit'' $\epsilon:=\frac{\ell}{L}\to 0$ scales the perforated domain such that perforations
become invisible in the macroscale.}
\label{fig:MicMac}
\end{center}
\end{figure}
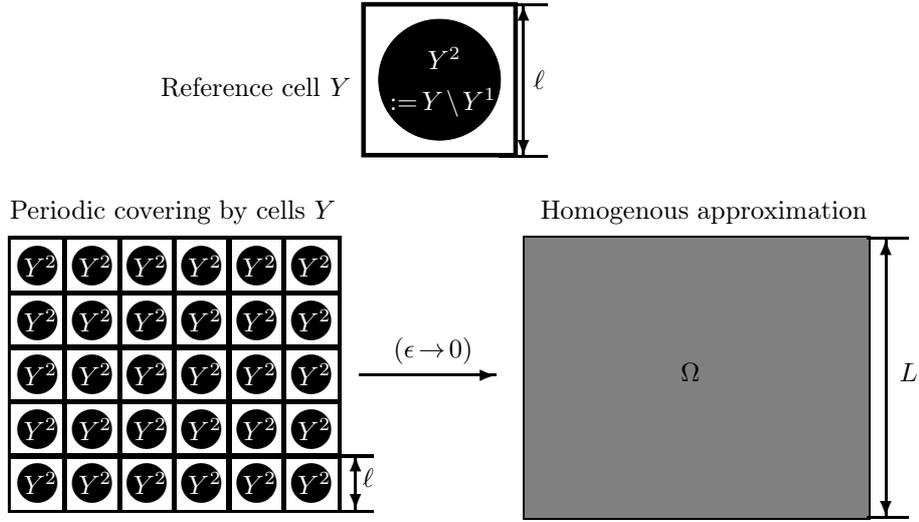

Here, we study the energy density
\reff{FrEn} with respect to a perforated domain $\Omega^\epsilon\subset\mathbb{R}^d$
instead of a homogeneous $\Omega\subset\mathbb{R}^d$. The dimensionless variable $\epsilon>0$ defines 
the heterogeneity $\epsilon =\frac{\ell}{L}$ where $\ell$ represents
the characteristic pore size and $L$ is the characteristic length of the porous medium, see Figure \ref{fig:MicMac}. Hence, 
the porous medium is characterized by a reference cell $Y:= [0,\ell_1]\times[0,\ell_2]\times\dots\times[0,\ell_d]$
which represents a single, characteristic pore. For simplicity, we set $\ell_1=\ell_2=\dots=\ell_d=1$.  A well-accepted approximation is then the periodic covering of the
macroscopic porous medium by such a single reference cell $\epsilon
Y$, see Figure \ref{fig:MicMac}. The pore
and the solid phase
of the medium are denoted by $\Omega^\epsilon$ and $B^\epsilon$, respectively.
These sets are defined by,
\bsplitl{
\Omega^\epsilon
    & := \bigcup_{{\bf z}\in\mathbb{Z}^d}\epsilon\brkts{Y^1+{\bf z}}\cap\Omega\,,
\qquad
B^\epsilon
    := \bigcup_{{\bf z}\in\mathbb{Z}^d}\epsilon\brkts{Y^2+{\bf z}}\cap\Omega
    =\Omega\setminus\Omega^\epsilon\,,
}{Oe2}
where the subsets $Y^1,\,Y^2\subset Y$ are defined such that $\Omega^\epsilon$
is a connected set. More precisely, $Y^1$ stands for the pore phase (e.g. liquid or gas phase in wetting problems),
see Figure \ref{fig:MicMac}.

These definitions allow us to reformulate \reff{PhMo}
by the following microscopic porous media problem
\bsplitl{
\textrm{(Micro porous case)}\,\,\,
\begin{cases}
\quad \partial_t\phi_\epsilon
    = {\rm div}\brkts{
    \hat{\rm M}\nabla \brkts{
            -\lambda^2\Delta \phi_\epsilon
            +f(\phi_\epsilon)
        }
    }
    & \quad\textrm{in }\Omega^\epsilon_T\,,
\\\quad
\nabla_n\phi_\epsilon:= {\bf n}\cdot\nabla\phi_\epsilon
    = 0
    & \quad\textrm{on }\partial\Omega^\epsilon_T
    \,,
\\\quad
\nabla_n\Delta\phi_\epsilon
    = 0
    & \quad\textrm{on }\partial\Omega^\epsilon_T
    \,,
\\\quad
\phi_\epsilon({\bf x},0)
    = \psi({\bf x})
    & \quad\textrm{on }\Omega^\epsilon\,.
\end{cases}
}{PeMoPr}
In the next section, we motivate our main goal of deriving a homogenized upscaled problem by passing
to the limit $\epsilon\to 0$ in \reff{PeMoPr}.

\subsection{Physical motivation}\label{sec:PhMo}

There is a large amount of literature available on multiphase flow
through porous media: e.g.~the review by~\cite{RevModPhys.65.1393}
on fluid flow in reservoir rocks and references therein, the
experimental works on viscous fluid imbibition processes in a
Hele-Shaw cell by~\cite{PhysRevLett.63.1685,hernandez_m_et_al_01},
\cite{PhysRevLett.89.104503} and \cite{PhysRevE.76.056312}, or the
study of fluid flow in sheets of paper
in~\cite{PhysRevLett.90.096101}, to name a few). 
A physically complex problem of vapor sorption and desorption from 
nanoporous solids is studied by \cite{Bazant2011}.
\cite{Adler1988}
provide a comprehensive review of the field. In this study, the
authors outline in detail some of the fundamental concepts of the
field, such as volume averaging and extending Darcy's law towards
two-phase flows, both used often up to date. Notably, the volume
averaging method requires a fictitious length scale defining the
test volumes. These volumes cannot be chosen to be the
characteristic pore scale as in homogenization theory in order to
comply with the ergodic hypothesis required by the method. The above
review also addresses the frequently questioned approach of using
phenomenological relative permeabilities \cite[]{Muskat1936}.

Rencently, \cite{Papatzacos2002,Papa2010} applied a special type of
volume averaging, Marle's averaging technique \cite[]{Marle1982}, to
a coupled system consisting of the continuity equation, a momentum
and an energy balance. The effective model then turns via Darcy's
law into a Cahn-Hilliard type equation for a phenomenologically
motivated transport parameter. This thermodynamic derivation of an
effective macroscopic Cahn-Hilliard equation for mass transport
starting from a microscopic continuity equation clearly demonstrates
the relevance of phase field type approaches in heterogeneous
structures. In fact, the use of the Cahn-Hilliard equation 
to describe macroscopic fluid flows in porous media has rececived a
lot of attention over the last few years. It has been shown that
such a phase field model adapted to imbibition reduces to 
Darcy's law in the sharp interface limit, i.e., when $\lambda\to 0$
(see e.g.~\cite{alava_04}). Therefore it is an ideal candidate,
particularly for numerical modelling, to study e.g.~the statistical
and dynamical properties of the kinetic roughening process that the
interface undergoes as  the liquid invades the porous
medium~\cite[]{Dube1999,hernandez_m_et_al_01,Laurila_etal_05,PhysRevE.74.041608}.

However, up to now, no effective macroscopic equations have been
derived for any microscopic porous media formulation \reff{PeMoPr}.
It should be noted that understanding rationally and systematically
how microscopic details affect global macroscopic properties is of a
crucial point in a wide spectrum of multiphase flows applications,
from traditional ones, such as oil recovery, to more recent ones,
such as micro- and nano-fluidics. The present study aims to address this
issue and at the same time exemplify its physical relevance for the
field of multiphase flows by using as a paradigm  the problem of
wetting in heterogeneous domains such as imbibition. As far as the
Cahn-Hilliard equation is concerned, it has a long history and
enjoys a broad range of applicability as discussed below. This is a
major motivation for the first homogenization result derived here in
the context of perforated or strongly heterogeneous domains.
Moreover, the upscaled problem should allow for efficient and
systematic low-dimensional computations in applications.

\subsection{On the broad applicability of the Cahn-Hilliard equation}\label{sec:ReLi}
As noted above, the Cahn-Hilliard equation has a wide applicability.
The phase-field equation \reff{PhMo} was first introduced by
\cite{Cahn1958} where they suggested a free-energy formulation for
nonuniform systems. Alternatively, Cahn-Hilliard-type equations can
be obtained by square-gradient approximations to non-local
free-energy functionals like those used in the statistical mechanics
of non-homogeneous fluids (e.g.~\cite{Miranville2003,Pereira2012}).
Since the work of Cahn and Hilliard, this formalism has become a
fundamental modeling tool in both science and engineering.
Cahn-Hilliard or more generally phase-field energy functionals are
for example applied in image processing such as inpainting, see
e.g.~\cite{Bertozzi2007}.
Wetting phenomena, of great interest in technological applications,
especially motivated by recent developments in micro-fluidics, enjoy
a wide-spread use of phase-field
modeling~(e.g.~\cite{Pomeau2001,Pradas2008,Pradas2011}). Such
phenomena have some intriguing features, including the appearance of
hysteresis and non-locality, e.g. correlations between the contact
line dynamics at each surface plate of a
micro-channel~\cite[]{Wylock2012}. Additional complexities in
wetting include the presence of an electric field (electrowetting,
e.g.~\cite{Eck2009}). There are numerous other applications where
phase-field models provide a powerful modelling tool. For example,
in \cite{Lowengrub2009} a phase-field model is proposed to describe
the dynamics of vesicles and associated phenomena, such as spinodal
decomposition, coarsening, budding, and fission. In this study, in
addition to the Cahn-Hilliard equation an Allen-Cahn equation
($L^2$-gradient flow of $E(\phi)$) is employed.

Clearly, there is a large amount of literature on
phase-field/Cahn-Hilliard models on a wide variety of physical
settings and applications, which cannot be fully reviewed here. That
said, it is important to emphasize that the key to the versatility
of phase-field/Cahn-Hilliard formulations is precisely the fact that
many physical settings are characterised  by simple energies of the
form \reff{FrEn}.

\medskip

In Section \ref{sec:2Fo} we introduce two relevant formulations
of the Cahn-Hilliard equation. The main theorem, which states the
new macroscopic Cahn-Hilliard equation, is given in Section
\ref{sec:MaRe}, where we also provide the local equilibrium
condition required for homogenization. In Section \ref{sec:Pr} we
demonstrate the applicability of the new effective equation in the
context of wetting and are able to connect it to physically
suggested models in imbibition. Conclusions and suggestions for
further work are presented in the Section~\ref{sec:Concl}.

\section{Two reformulations of the Cahn-Hilliard equation: Zero mass and splitting}\label{sec:2Fo}
We present two equivalent formulations of the Cahn-Hilliard
equation. The first helps to achieve solvability for Lipschitz
inhomogeneities and the second, referred to as ``splitting
formulation'', decouples the Cahn-Hilliard equation into two second
order problems for a feasible upscaling by the multiple-scale
method.

\medskip

{\bf (i) Zero mass formulation (for well-posedness)}\label{sec:Ex}
\cite{Novick-Cohen1990} proves well-posedness of the Cahn-Hilliard
problem \reff{PhMo} rewritten for $\Omega_T:=\Omega\times]0,T[$ and $\partial\Omega_T:=\partial\Omega\times]0,T[$ in the following zero mass formulation
\bsplitl{
\textrm{\bf (Zero mass)}\quad
\begin{cases}
\quad
\partial_t v
    = {\rm div}\brkts{\hat{\rm M}\nabla\brkts{
        b v + h(v) -\lambda^2\Delta v
    }
    }
    & \quad\textrm{in }\Omega_T\,,
\\
\quad
\nabla_n v
    = {\bf n}\cdot\nabla\Delta v
    =0
    & \quad\textrm{on }\partial\Omega_T\,,
\\
\quad
v({\bf x},0)
    = v_0({\bf x})
    = \psi({\bf x})-\ol{\phi}\,
    &\quad\textrm{in } \Omega,
\end{cases}
}{ZeMaFo}
where $v({\bf x},t):=\phi({\bf x},t)-\ol{\phi}$, $b:=f'(\phi)$, $h(v):=f(\ol{\phi}+v)-bv$, and by mass conservation of
\reff{PhMo} we define
$\frac{1}{\av{\Omega}}\int_\Omega \phi\,d{\bf x}
    :=
    \frac{1}{\av{\Omega}}\int_\Omega \psi\,d{\bf x}
    =: \ol{\phi}\,.
$ 
These definitions imply $bv+h(v)=f(\ol{\phi}+v)$.
For $k\geq 0$, we introduce the family of spaces
\bsplitl{
H^k_E(\Omega)
    = \brcs{
        \phi\in H^k(\Omega)\,\Bigr|\,\nabla_n\phi=0\textrm{ and }\ol{\phi}=0
    }\,.
}{EnSp}
\cite{Novick-Cohen1990} verifies local existence and uniqueness of solutions
$v\in H^2_E(\Omega)$ of problem \reff{ZeMaFo} for
$f\in C^2_{Lip}(\mathbb{R})$ with $\av{f(s)}\to\infty$ as $s\to\pm\infty$ and
$v({\bf x},0)\in H^2_E(\Omega)$. Moreover, in \cite{Novick-Cohen1990} one also finds necessary
conditions on $h$ leading to global existence.

\medskip
{\bf (ii) Splitting (for homogenization)}\label{sec:Sp}
The existence result summarized in the previous section enables us to give the following
weak formulation of problem \reff{PhMo}. There exists for all $\varphi\in H^2_E(\Omega)$ a
weak solution $v\in H^2_E(\Omega)$ solving the equation
\bsplitl{
\frac{d}{dt}(v,\varphi)
    + \lambda^2\brkts{
            \Delta v
        ,{\rm div}\brkts{{\hat{\rm M}}\nabla \varphi }
    }
    =\brkts{{\rm div}\brkts{{\hat{\rm M}}\nabla f(\ol{\phi}+v)}
        ,\varphi}\,.
}{WeSpFo}
By identifying
$v
    = (-\Delta)^{-1}w$
in the $H^2_E(\Omega)$-sense together with solvability of equation \reff{WeSpFo}
we are able to introduce
the following problem
\bsplitl{
\textrm{\bf (Splitting)}\quad
\begin{cases}
\quad \partial_t(-\Delta)^{-1}w
    -\lambda^2{\rm div}\brkts{
    \hat{\rm M}\nabla
        w
    }
        = {\rm div}\brkts{
    \hat{\rm M}\nabla
        f(\ol{\phi}+v) 
    }
    & \textrm{in }\Omega_T\,,
\\\quad
\nabla_n w
    = -\nabla_n\Delta v
    = 0
    &\textrm{on }\partial\Omega_T\,,
\\\quad
-\Delta v
    = w
    &\textrm{in }\Omega_T\,,
\\\quad
\nabla_n v
    = g({\bf x})
    &\textrm{on }\partial\Omega_T\,,
\\\quad
v({\bf x},0)
    = \psi({\bf x})
    -\ol{\phi}
    & \textrm{in }\Omega\,,
\end{cases}
}{DePhMo0}
which is equivalent to \reff{PhMo} in the $H^2_E$-sense and hence, when $g=0$, is well-posed too, \cite{Novick-Cohen1990}.
The advantage of \reff{DePhMo0} is that it allows to base our upscaling approach
on well-known results from elliptic/parabolic homogenization theory \cite[]{Bensoussans1978,Pavliotis2008,Zhikov1994}.
Finally, we remark that the splitting \reff{DePhMo0} slightly differs from the strategy applied for computational
purposes in \cite{Barrett1999}, for instance.

\section{Main results}\label{sec:MaRe}

Before we state our main result, the subsequent homogenization of the Cahn-Hilliard equation
requires the assumption of local thermodynamic equilibrium.

\begin{defn}\label{def:LoEq} {\em (Local equilibrium)}
We say that the phase-field $\phi$ is in local thermodynamic equilibrium, 
if and only if
\bsplitl{
\frac{\delta E(\phi)}{\delta \phi}
    = \mu(\phi)
    = f(\phi)
    -\lambda^2\Delta \phi
    =
    {\rm const.}\,,
}{LoEqCo}
for each ${\bf x}/\epsilon={\bf y}$ element of the same reference cell $Y$. $\mu$ stands for the chemical potential which is only allowed to vary over the different
reference cells.
\end{defn}

The  state of general conditions of equilibrium of heterogeneous
substances seems to go back to the celebrated work of
\cite{Gibbs1876}. The assumption of local thermodynamic equilibrium
can be justified on physical and mathematical grounds by the assumed
separation of macroscopic (size of the porous medium) and
microscopic (characteristic pore size) length scales and the
emerging difference in the associated characteristic timescales.
This kind of equilibrium assumptions are widely applied to a veriety
of physical situations such as diffusion \cite[]{Nelson1999},
macroscale thermodynamics in porous media \cite[]{Benn1999} and
ionic transport in porous media based on dilute solution theory
\cite[]{Schmuck2011a,Schmuck2012,Schmuck2012a}, for instance. Local equilibrium
assumptions as in Definition \ref{def:LoEq} emerge as key
requirements for the mathematical well-posedness of arising cell
problems which define effective transport coefficients in
homogenized, nonlinear (and coupled) problems.

The homogeneous free energy $F$ in \reff{hFrEn} enables the upscaling 
under the following

\medskip

{\bf Assumption F:} \emph{Assume that the homogeneous free energy $F$ satisfies for 
real parameters $\alpha_2>\alpha_1>0$, which define $F$ as a double-well potential by 
$F(s) = (s-\alpha_1)^2(s-\alpha_2)^2\,,$
such that
\bsplitl{
25(\alpha_1+\alpha_2)^2-20(\alpha_1^2+\alpha_2^2+3\alpha_1\alpha_2)
	> (\alpha_1+\alpha_2)^2/4\,.
}{PC}
}

These considerations allow us to state the following main result of this study.
\begin{thm}\label{thm:EfPhFi}\emph{(Upscaled Cahn-Hilliard equations)}
Let $\hat{\rm M}=\brcs{{\rm m}\delta_{ij}}_{1\leq i,j\leq d}$ for ${\rm m}>0$ be
an isotropic mobility tensor. We assume that the local equilibrium 
condition \reff{LoEqCo} is satisfied.
Moreover, suppose that $\psi({\bf x})\in H^2_E(\Omega)$ and let $F$ satisfy Assumption {\bf F}.
Then, the microscopic porous media formulation \reff{PeMoPr} can be effectively
approximated by the following macroscopic problem,
\bsplitl{
\begin{cases}
\theta_1\pd{\phi_0}{t}
    = {\rm div}\biggl(
        \Bigl[
            \theta_1f'(\phi_0)\hat{\rm M}
            -\Bigl( 2\frac{f(\phi_0)}{\phi_0}
            - f'(\phi_0)
            \Bigr)\hat{\rm M}_v
        \Bigr]\nabla \phi_0
    \biggr)
\\
\qquad
    -f'(\phi_0){\rm div}\brkts{
        \hat{\rm M}_v\nabla \phi_0
    }
    +\frac{\lambda^2}{\theta_1}{\rm div}\brkts{
        \hat{\rm M}_w\nabla \brkts{
            {\rm div}\brkts{
                \hat{\rm D}\nabla \phi_0
            }
        }
    }
    &\textrm{in }\Omega_T\,,
\\
\nabla_n \phi_0
    = {\bf n}\cdot\nabla\phi_0
    = 0
    &\textrm{on }\partial\Omega_T\,,
\\
\nabla_n\Delta \phi_0
    = 0
    &\textrm{on }\partial\Omega_T\,,
\\
\phi_0({\bf x},0)
    = \psi({\bf x})
    &\textrm{in }\Omega\,,
\end{cases}
}{pmWrThm}
where $\theta_1:=\frac{\av{Y^1}}{\av{Y}}$ is the porosity and the porous media correction tensors $\hat{\rm D}:=\brcs{{\rm d}_{ik}}_{1\leq i,k\leq d}$,
$\hat{\rm M}_v = \brcs{{\rm m}^v_{ik}}_{1\leq i,k\leq d}$ and
$\hat{\rm M}_w=\brcs{{\rm m}^w_{ik}({\bf x})}_{1\leq i,k\leq d}$
 are defined by
\bsplitl{
{\rm d}_{ik}
    & := \frac{1}{\av{Y}}\sum^d_{j=1}\int_{Y^1}\brkts{
        \delta_{ik} - \delta_{ij}\pd{\xi^k_v}{y_j}
        }
    \,d{\bf y}\,,
\\
{\rm m}^v_{ik}
    & :=
    \frac{1}{\av{Y}}\sum_{j=1}^d\int_{Y^1}{\rm m}\brkts{
        \delta_{ik}
        -\delta_{ij}\pd{\xi^k_v}{y_j}
    }\,d{\bf y}\,,
\\
{\rm m}^w_{ik}({\bf x})
    & :=
    \frac{1}{\av{Y}}\sum_{j=1}^d\int_{Y^1}{\rm m}\brkts{
        \delta_{ik}
        -\delta_{ij}\pd{\xi^k_w({\bf x})}{y_j}
    }\,d{\bf y}\,.
}{Dik}
The corrector functions $\xi^k_v\in H^1_{per}(Y^1)$ and $\xi^k_w\in L^2(\Omega;H^1_{per}(Y^1))$ for $1\leq k\leq d$
solve in the distributional sense the 
following reference cell problems
\bsplitl{
\xi_w^k:\quad
\begin{cases}
-\sum_{i,j,k=1}^d
    \pd{}{y_i}\brkts{
        \delta_{ik}-\delta_{ij}\pd{\xi^k_w}{y_j}
    }
\\\qquad\qquad
    =
    \lambda^2 \sum_{k,i,j=1}^d\pd{}{y_i}\brkts{
        {\rm m}_{ik}
        -\frac{f(\phi_0)}{f'(\phi_0)\phi_0}{\rm m}_{ij}\pd{\xi^k_v}{y_j}
    }
    &\textrm{ in }Y^1\,,
\\
\sum_{i,j,k=1}^d{\rm n}_i
        \brkts{
        \delta_{ij}\pd{\xi^k_w}{y_j}
        -\delta_{ik}
        }
\\\qquad\qquad
        -\lambda^2 \sum_{k,i,j=1}^d\pd{}{y_i}\brkts{
            {\rm m}_{ik}
            -\frac{f(\phi_0)}{f'(\phi_0)\phi_0}{\rm m}_{ij}\pd{\xi^k_v}{y_j}
        }
    \Bigr)
    = 0
    &\textrm{ on }\partial Y^1\,,
\\
\xi^k_w({\bf y})\textrm{ is $Y$-periodic and ${\cg M}_{Y^1}(\xi^k_w)=0$,}
\end{cases}
\\
\xi_v^k:\quad
\begin{cases}
-\sum_{i,j=1}^d
    \pd{}{y_i}\brkts{
        \delta_{ik}-\delta_{ij}\pd{\xi^k_v}{y_j}
    }
    = 0
    &\textrm{ in }Y^1\,,
\\
\sum_{i,j=1}^d{\rm n}_i
	\brkts{
        \delta_{ij}\pd{\xi^k_w}{y_j}
        -\delta_{ik}
        }
    =
    0
    &\textrm{ on }\partial Y^1\,,
\\
\xi^k_v({\bf y})\textrm{ is $Y$-periodic and ${\cg M}_{Y^1}(\xi^k_v)=0$.}
\end{cases}
}{pmRCTh}
\end{thm}

\begin{rem}\label{rem:Thm}
i) The reference cell problem \reff{pmRCTh}$_1$ for $\xi^k_v$ can be solved numerically for
example. For problem \reff{pmRCTh}$_2$, there are results in the literature [e. g. \cite{Auriault1997}] in 
the case of straight or perturbed straight channels. \\
ii) The thermodynamic equilibrium \reff{LoEqCo} enables the derivation of the 
cell problem \reff{pmRCTh}$_1$ and Assumption {\bf F} is necessary for its well-posedness. \\
\end{rem}

\section{Applications to wetting}\label{sec:Pr}
 The freedom in defining the free energy $F(\phi)$ in the phase field
equation~\reff{PhMo}, enables us to apply the upscaling formalism
developed in this paper to a variety of physical problems.  
Taking $F$ as in Assumption {\bf F} includes the phenomenological double-well form 
which is generally applied for the homogeneous free energy. Herewith, we can immediately describe the evolution
of two phases such as liquid--gas through a porous medium for
instance. The quantity of interest in describing wetting phenomena
is the contact angle, defined as the angle between the liquid--gas
interface and the wetted area of the substrate.

In the phase field model \reff{PhMo} it is well accepted to account
for wetting properties by a Robin boundary condition \reff{PhMo}$_2$
(e.g.~\cite{Wylock2012}) with \bsplitl{ g({\bf x})
    := -\frac{\gamma}{C_h}a({\bf x})\,.
}{RoBC}
The parameter $C_h$ is the Cahn number $\lambda/L$ and
$\gamma=2\sqrt{2}\phi_e/3\sigma_{lg}$ where $\sigma_{lg}$ denotes the liquid-gas surface tension and
$\phi_e$ the local equilibrium limiting values of $F$. It is straightforward to extend \reff{RoBC} to several wetting properties $a_1,\,a_2,\,\dots,a_N$ for a positive
$N\in\mathbb{N}$ such that
\bsplitl{
g({\bf x})
    := -\frac{\gamma}{C_h}\sum_{i=1}^N a_i({\bf x})\chi_{\partial\Omega_w^i}({\bf x})
    \quad\in H^{3/2}(\partial\Omega_w)\,.
}{aN}
For notational brevity, we will work with $N=2$ in subsequent sections.

In the Conclusion (Section~\ref{sec:Concl}) we briefly relate the results obtained in this paper to the results
from~\cite{Alberti2005} where a formula for the effective contact
angle is derived based on $\Gamma$-convergence and geometric measure
theory.

\subsection{Channel with heterogeneous wetting properties}\label{sec:ChHeWe}
\begin{figure}[htbp]
\begin{center}


\setlength{\unitlength}{.8cm}
\begin{picture}(8,6)(0,0)


\begin{tikzpicture}[decoration=brace,scale=2.5, line width=1pt] 

    \coordinate (A1) at (0, 0);
    \coordinate (A2) at (0, 1);
    \coordinate (A3) at (1, 1);
    \coordinate (A4) at (1, 0);
    \coordinate (B1) at (0.3, 0.3);
    \coordinate (B2) at (0.3, 1.3);
    \coordinate (B3) at (1.3, 1.3);
    \coordinate (B4) at (1.3, 0.3);


%

    \draw[very thick] (A1) -- (A2);
    \draw[very thick] (A2) -- (A3);
    \draw[very thick] (A3) -- (A4);
    \draw[very thick] (A4) -- (A1);

    \draw[dashed] (A1) -- (B1);
    \draw[dashed] (B1) -- (B2);
    \draw[very thick] (A2) -- (B2);
    \draw[very thick] (B2) -- (B3);
    \draw[very thick] (A3) -- (B3);
    \draw[very thick] (A4) -- (B4);
    \draw[very thick] (B4) -- (B3);
    \draw[dashed] (B1) -- (B4);

    \draw[thick] (1.3,0.3) -- (1.5,0.3);
    \draw[thick] (1.0,0.0) -- (1.2,0.0);
    \draw[very thick,<->] (1.1,0.0) -- (1.4,0.3);
    \begin{scope}[every node]
       \node [font=\large, color=black] at (1.35, 0.15) {$1$};
    \end{scope}

    \draw[thick] (0.0,-0.2) -- (0.0,0.0);
    \draw[thick] (1.0,-0.2) -- (1.0,0.0);
    \draw[very thick,<->] (0.0,-0.1) -- (1.0,-0.1);
    \begin{scope}[every node]
       \node [font=\large, color=black] at (0.5, -0.2) {$1$};  
    \end{scope}

    \draw[thick] (1.3,1.3) -- (1.5,1.3);
    \draw[very thick,<->] (1.4,0.3) -- (1.4,1.3);
    \begin{scope}[every node]
       \node [font=\large, color=black] at (1.5, 0.8) {$1$};
    \end{scope}

   \begin{scope}[every node]
    \node [font=\Huge, color=black] at (-0.4,0.5) {$Y:=$};
   \end{scope}

   \begin{scope}[every node/.append style={xslant=0.3},xslant=1.0]
      \node [font=\large, color=gray] at (-0.45, 1.13) {$\pmb{\partial Y_{w_2}}$};
   \end{scope}

   \begin{scope}[every node]
    \node [font=\large, color=gray] at (0.55,0.5) {$\pmb{\partial Y_{w_1}}$};
   \end{scope}

   \begin{scope}[every node]
    \node [font=\large, color=gray] at (0.8,0.8) {$\pmb{\partial Y_{w_4}}$};
   \end{scope}

   \begin{scope}[every node/.append style={yslant=1.3},yslant=1]
    \node [font=\large, color=gray] at (1.155,-0.5) {$\pmb{\partial Y_{r}}$};
   \end{scope}

   \begin{scope}[every node/.append style={yslant=1.3},yslant=1]
    \node [font=\large, color=gray] at (0.15,0.5) {$\pmb{\partial Y_{l}}$};
   \end{scope}

   \begin{scope}[every node/.append style={xslant=0.3},xslant=1.0]
      \node [font=\large, color=gray] at (0.5, 0.13) {$\pmb{\partial Y_{w_3}}$};
   \end{scope}

\end{tikzpicture}
\end{picture}
\caption{Reference channel $Y:=[0,1]^d$ defined by channel entry $\partial Y_l$, channel exit $\partial Y_r$,  and wall $\partial Y_w:=\bigcup_{i=1}^4\partial Y_{w_i}$ where $\partial Y_{w_i}:=\partial Y_{w_i}^1\cup\partial Y^2_{w_i}$ with two
different wetting properties $\partial Y^1_{w_i}$ and $\partial Y^2_{w_i}$. We point out that $Y$ is only scaled
in $y_1=\frac{x_1}{\epsilon}$ direction and keeps $y_2$ and $y_3$ fixed.
}
\end{center}
\label{fig:YCh}
\end{figure}
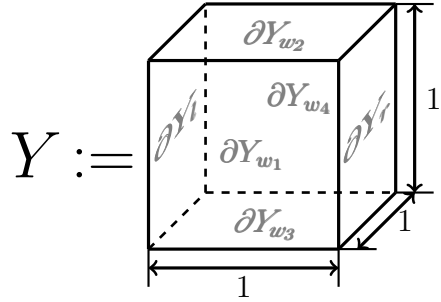

We assume that $\Omega:=[0,L]\times [0,1]^{d-1}\subset\mathbb{R}^d$ is an arbitrary straight channel of length $L$ with walls $\partial\Omega_w$
having different wetting properties. We assume that these wetting properties
repeat periodically along the channel walls. We denote the left entrance by $\Gamma^l$ and the right exit by $\Gamma^r$ 
such that $\partial\Omega=\Gamma^l\cup\partial\Omega_w\cup\Gamma^r$. In particular, we define
\bsplitl{
\Omega^\epsilon
     := \brcs{\bigcup_{z\in\mathbb{Z}} {\bf e}_{\epsilon}\brkts{Y+ z{\bf e}_1}}\cap\Omega
     &\,,
\qquad
\partial\Omega_w^\epsilon
    := \brcs{\bigcup_{z\in\mathbb{Z}}{\bf e}_{\epsilon}\brkts{ Y+ z{\bf e}_1}}\cap\partial\Omega_w\,,
}{HCh}
where ${\bf e}_\epsilon=\epsilon{\bf e}_1+{\bf e}_2+\dots+{\bf e}_d$, ${\bf e}_i$ for $i=1,2,\dots,d$ is the canonical basis of $\mathbb{R}^d$, and for the definition of $Y$ we refer to Figure \ref{fig:YCh}.

To derive an effective phase-field model for highly heterogeneous
walls, we account for different surface properties on the walls
$\partial\Omega_w^\epsilon$, see \reff{HCh}, by the following multiscale
formulation, \bsplitl{
\begin{cases}
\quad \partial_t(-\Delta)^{-1}w_\epsilon
    = {\rm div}\brkts{
    \hat{\rm M}\nabla \brkts{
            \lambda^2w_\epsilon
            -\phi_\epsilon
            +\phi^3_\epsilon
        }
    }
    & \textrm{in }\Omega_T^\epsilon\,,
\\\quad
-\Delta \phi_\epsilon
    = w_\epsilon
    &\textrm{in }\Omega_T^\epsilon\,,
\\\quad
{\bf n}\cdot{\bf J}_\epsilon
    = J_l
    &\textrm{on }\Gamma^l_T:=\Gamma^l\times ]0,T[\,,
\\\quad
{\bf n}\cdot {\bf J}_\epsilon
    = 0
    &\textrm{on }\Gamma^r_T:=\Gamma^r\times ]0,T[\,,
\\\quad
\nabla_n\phi_\epsilon
    = -\epsilon
    g({\bf x}/\epsilon)
    &\textrm{on }\partial\Omega_w^\epsilon\times ]0,T[\,,
\\\quad
\phi_\epsilon({\bf x},0)
    =\psi({\bf x})
    &\textrm{in }\Omega\,,
\end{cases}
}{DePhMo}
where ${\bf J}_\epsilon$ is defined as the flux
$\nabla \brkts{\lambda^2w_\epsilon-\phi_\epsilon+\phi^3_\epsilon}$, and $a_1$ and $a_2$
are constants.

For the homogenization of heterogeneous boundary conditions such as \reff{DePhMo}$_5$, we
refer to \cite{Allaire1996}. Problem \reff{DePhMo} is introduced because it is \emph{a priori} not clear whether
oscillations on the solid/void interface, i.e., on the walls $\partial\Omega_w^\epsilon$, also influence the bulk. We also
need to properly
define the periodic microscale $x_1/\epsilon=:y_1$. We assume that the heterogeneities
defined on the wall $\partial\Omega_{w}^\epsilon$ are periodic
in the $x_1$-direction with period defined via a reference cell as in Figure \ref{fig:YCh}. Our
averaging process consists in the usual limit $\epsilon\to 0$. Hence, we cover the channel $\Omega$ by
reference cells $Y$, e.g. as in Figure \ref{fig:YCh}, which are only scaled by $\epsilon$
in the $x_1$-direction. We further need the following:

\medskip
{\bf Hypothesis (HI):} \emph{We assume that the boundary $\partial\Omega_w$ contains
finitely many flat pieces with conormal not proportional to any ${\bf z}\in\mathbb{Z}^d$.}

\medskip
If the Hypothesis {\bf (HI)} is violated, then the homogenization limit does not converge
towards a unique upscaled problem, see \cite{Bensoussans1978}.
%
%

\begin{cor}\label{thm:EfWa}\emph{(Heterogeneous walls)} We make the same assumptions as in Theorem
\ref{thm:EfPhFi} except that we do not require an isotropic mobility $\hat{\rm M}$. We additionally suppose
that {\bf (HI)} holds and that $J_l,\,g\in H^{3/2}(\partial\Omega)$ in \reff{DePhMo}.

Then, the microscopic wall description \reff{DePhMo} becomes the following upscaled system after averaging
over the microscale,
\bsplitl{
\begin{cases}
\partial_t\phi_0
    = {\rm div}\brkts{
        \hat{\rm M}\nabla \brkts{
            -\phi_0
            +\phi^3_0
            -\lambda^2\Delta\phi_0
        }
    }
    &\textrm{in }\Omega_T\,,
\\
\nabla_n \phi_0
    = J_l
    &\textrm{on }\partial\Omega^l_T\,,
\\
\nabla_n \phi_0
    = 0
    &\textrm{on }\partial\Omega^r_T\,,
\\
\nabla_n \phi_0
    = g_0
    &\textrm{on }
        \partial\Omega_w\times ]0,T[\,,
\\
\phi({\bf x},0)
    =\psi({\bf x})
    &\textrm{in }\Omega\,,
\end{cases}
}{EfWr}
where
$g_0
    := -\frac{\gamma}{C_h}\frac{1}{\av{Y}}\int_Y
        \brkts{
            a_1\chi_{\partial Y^{1}_{w}}({\bf y})
            +a_2\chi_{\partial Y^{2}_{w}}({\bf y})
        }
    \,d{\bf y}
    \,,$
where the constants $a_1$ and $a_2$ characterize the material's wetting properties.
\end{cor}
%
%
%

\subsection{Wetting dynamics in porous media and imbibition}\label{sec:WePm}
As in equation \reff{Oe2}, 
we define the porous medium by the
pore space $\Omega^\epsilon$ and the solid material $B^\epsilon$ as a periodic covering
by a single reference cell $Y:= [0,\ell_1]\times[0,\ell_2]\times\dots\times[0,\ell_d]$ which defines
the characteristic pore geometry, see Figure \ref{fig:MicMac}.


We denote by
$\partial Y^1_w:=\bigcup_{i=1}^N\partial Y^1_{w_i}$ the pore surface. The subsets $\partial Y^1_{w_i}$ belong to surfaces
with different wetting properties. Correspondingly, the walls $\partial\Omega^\epsilon_{w_i}$
are defined via $\partial Y^1_{w_i}$ of the covering of $\Omega$ by $Y$, see Figure \ref{fig:YCh}. Depending on applications, different boundary conditions than \reff{PeMo2}$_4$ below
for wetting can be imposed.

These definitions allow to reformulate \reff{PhMo}
by the following microscopic porous media problem,
\bsplitl{
\begin{cases}
\quad \partial_t\phi_\epsilon
    = {\rm div}\brkts{
    \hat{\rm M}\nabla \brkts{
            -\lambda^2\Delta \phi_\epsilon 
            +f(\phi_\epsilon)
        }
    }
    & \textrm{in }\Omega^\epsilon_T\,,
\\\quad
{\bf n}\cdot{\bf J}_\epsilon
    = J_l
    &\textrm{on }\Gamma^l_T\,
\\\quad
{\bf n}\cdot {\bf J}_\epsilon
    = 0
    &\textrm{on }\Gamma^r_T\, 
\\\quad
\nabla_n\phi_\epsilon
    = -\epsilon\frac{\gamma}{C_h}\brkts{
        a_1({\bf x})\chi_{\partial\Omega^\epsilon_{w_1}}({\bf x}/\epsilon)
        +a_2({\bf x})\chi_{\partial\Omega^\epsilon_{w_2}}({\bf x}/\epsilon)
    }
    &\textrm{on }\partial\Omega^\epsilon_w\times ]0,T[\,,
\end{cases}
}{PeMo2}
where $a_1({\bf x})$ and $a_2({\bf x})$ appear periodically with period $\epsilon Y$ and vary macroscopically
in ${\bf x}\in\Omega^\epsilon$.
We complement
\reff{PeMo2} with arbitrary initial conditions
$\phi_\epsilon({\bf x},0)
    = \psi({\bf x})\in H^2_E(\Omega)\,.$

 We focus here on a porous medium with walls showing only two different
wetting properties, i.e., $N=2$. An extension to arbitrary $0<N<\infty$ is straightforward. We explained the scaling by $\epsilon$ of the wetting boundary condition \reff{PeMo2}$_4$
already in Section \ref{sec:Pr}~\reff{sec:ChHeWe}.

\begin{cor}\label{thm:EfPm}\emph{(Wetting in porous media)}
We make the same assumptions as in Theorem \ref{thm:EfPhFi}.

Then, the microscopic porous media formulation \reff{PeMo2} has the following leading
order asymptotic equation on the macroscale,
\bsplitl{
\begin{cases}
\theta_1\pd{\phi_0}{t}
    = {\rm div}\biggl(
        \Bigl[
            \theta_1f'(\phi_0)\hat{\rm M}
            -\Bigl( 2\frac{f(\phi_0)}{\phi_0}
            - f'(\phi_0)
            \Bigr)\hat{\rm M}_v
        \Bigr]\nabla \phi_0
    \biggr)
\\
\qquad
    -f'(\phi_0){\rm div}\brkts{
        \hat{\rm M}_v\nabla \phi_0
    }
    +\frac{\lambda^2}{\theta_1}{\rm div}\brkts{
        \hat{\rm M}_w\nabla \brkts{
            {\rm div}\brkts{
                \hat{\rm D}\nabla \phi_0
            }
            -\tilde{g}_0
        }
    }
    &\textrm{in }\Omega_T\,,
\\
{\bf n}\cdot {\bf J}
    = J_l
    & \textrm{on }\Gamma^l_T\,,
\\
{\bf n}\cdot {\bf J}
    = 0
    & \textrm{on }\Gamma^r_T\,,
\\
\nabla_n \phi_0
    = {\bf n}\cdot\nabla\phi_0
    = \nabla_n\Delta \phi_0
    = 0
    &\textrm{on }\partial\Omega_w\times ]0,T[\,,
\\
\phi_0({\bf x},0)
    = \psi({\bf x})
    &\textrm{in }\Omega\,,
\end{cases}
}{pmWrTh}
where $\theta_1:=\frac{\av{Y^1}}{\av{Y}}$ is the porosity, ${\bf J}$ the flux corresponding to \reff{pmWrTh}$_1$, and the porous media correction tensors $\hat{\rm D}:=\brcs{{\rm d}_{ik}}_{1\leq i,k\leq d}$,
$\hat{\rm M}_v = \brcs{{\rm m}^v_{ik}}_{1\leq i,k\leq d}$ and
$\hat{\rm M}_w=\brcs{{\rm m}^w_{ik}({\bf x})}_{1\leq i,k\leq d}$
 are defined in \reff{Dik}.
The function $\tilde{g}_0$ defines the upscaled wetting boundary condition
\bsplitl{
\tilde{g}_0({\bf x})
    := -\frac{\gamma}{C_h}\int_{\partial Y^1_w}\brkts{
        a_1({\bf x})\chi_{\partial Y^1_{w_1}}({\bf y})
        +a_2({\bf x})\chi_{\partial Y^1_{w_2}}({\bf y})
    }\,ds({\bf y})\,.
}{UpScWeBC}
\end{cor}



\section{Conclusion}\label{sec:Concl}
We have examined the problem of upscaling the Cahn-Hilliard equation
for perforated/strongly heterogeneous domains. An effective
macroscopic Cahn-Hilliard equation by homogenization for such
domains is derived rigorously for the first time.  It is often the
case heuristic averaging strategies such as volume averages and
Marle's method are applied. However, such approaches are rather
heuristic and it is not clear how to choose the size of reference
volume for the averaging, see Section~\ref{sec:Intr}~\ref{sec:PhMo}. The proof of
the main result obtained here is valid for free energies $F$ defined
by polynomials up to $4$-th order satisfying Assumption {\bf F}. Such polynomial free
energies include generically applied double-well potentials which
phenomenologically represent a large class of free energies
modeling two-phase problems (e.g. the free energy of mixing
\reff{ReSoMo}) and which mimic the Lennard-Jones potential via the
LMP theory \cite[]{Presutti2009}. However, they do not appear as a
mean field limit of an atomistic model. Moreover, the upscaling
process also provides naturally the basic algorithmic framework and
analytical tools for other choices of free
energies $F$.

The new effective Cahn-Hilliard formulation introduces an efficient
and low-dimensional numerical alternative over its microscopic
counterpart \reff{PeMoPr} and serves as a promising alternative for
multiphase problems, see Section~\ref{sec:Intr}~\ref{sec:PhMo}.
Moreover, it provides systematically effective transport
coefficients like diffusion and mobility (or permeability) tensors.
We further apply the new effective Cahn-Hilliard equation to wetting
problems in porous media and straight channels. It turns out that
the new formulation allows for a feasible computation of effective
contact angles in channels with strongly heterogeneous walls for
instance. Interestingly, we recover rigorously the same equation
which was suggested in \cite{Ala-Nissila2004,Dube1999} for
imbibition but based on physical arguments, suggesting that the new
equation is consistent with known physical laws.

It should also be mentioned that Corollary~\ref{thm:EfPm} allows for
a formal extension towards random porous media or random wetting
properties where $a_1({\bf x})$ and $a_2({\bf x})$ are spatially
homogeneous and stationary ergodic random variables for instance. In
the case of random media, one can introduce appropriate random
variables such as a random porosity $\theta_1$ or random wall
fractions
$\theta_{w_1}:=\frac{\av{\partial Y^1_{w_1}}}{\av{\partial Y^1_w}}$
 and 
$\theta_{w_2}
    :=1-\theta_1\,.\,$
Herewith, we can redefine $g_0$ in \reff{UpScWeBC} by
\bsplitl{ \alpha({\bf x})
    := -\frac{\gamma}{C_h}\brkts{
        a_1\theta_{w_1}({\bf x})
        +a_2\theta_{w_2}({\bf x})
    }\,,
}{alpha} where $\theta_{w_i}({\bf x})$ for $i=1,2$ are homogeneous
random fields characterizing the wall fractions and the periodicity
assumption can be replaced by a stationary ergodic setting
\cite[]{Bensoussans1978}. Equation \reff{alpha} motivates that
homogenization theory allows to reliably introduce and consistently
define the phenomenological variable $\alpha$ appearing in the
equation for imbibition in \cite{Ala-Nissila2004,Dube1999}. In fact,
we obtain rigorously that this variable $\alpha$ is connected with
the wetting boundary condition $g$ in \reff{PeMoPr}. However, we
remark that the above extensions are merely formal and require
careful analytical considerations in specific applications of
interest.

Moreover, the effective model \reff{EfWr} allows us to determine the averaged
contact angle via $g_0$ in \reff{EfWr} or \reff{UpScWeBC}. We can determine
via $\gamma:=\frac{2\sqrt{2}\phi_e}{3\sigma_{lg}}$ the parameter
$a_{\rm eff}
    = \frac{g_0C_h}{\gamma}\,,$
and $\phi_e$ denotes the local equilibrium limiting values 
of the standard the phenomenological double-well potential $F$. 
By defining $\phi_e=+1$ as the liquid phase and $\phi_e=-1$ as the gaseous
phase, one imposes with $a_{\rm eff}>0$ hydrophilic and with $a_{\rm eff}<0$ hydrophobic wetting
conditions.
After setting $A=\sqrt{2}\gamma a_{\rm eff}$, the effective
equilibrium contact angle immediately follows by
\bsplitl{
{\rm cos}\,\theta_e
    = \frac{1}{2}\ebrkts{
        (1+A)^{3/2}
        -(1-A)^{3/2}
    }
    \,.
}{EA} We believe that herewith we can propose a convenient and
feasible alternative to \cite{Alberti2005} with \reff{EA} for the
computation of effective contact angles. Formula \reff{UpScWeBC},
allows to analytically compute the effective macroscopic contact
angle $\theta_e$ in contrast to the not easily accessible formulas
in \cite{Alberti2005}. The difference between theirs 
and our result relies on the fact that they 
work with
the interfacial energy \bsplitl{ E:=\sigma_{SL}\av{\Sigma_{SL}}
    +\sigma_{SV}\av{\Sigma_{SV}}
    +\sigma_{LV}\av{\Sigma_{LV}}
    + {\rm a.t.}\,,
}{IntfEn}
where $\sigma_{AB}$ denotes the surface tension between phases $A$ and $B$, $\Sigma_{AB}$ the
interface between $A$ and $B$ ($\av{\Sigma_{AB}}$ its measure), for $A,B\in \brcs{S,L,V}$. The letters $S,L$, and $V$ stand for the solid, liquid, and vapor phase, respectively. In contrast,
we base our considerations on the Cahn-Hilliard model \reff{DePhMo} and hence provide
an approximate effective contact angle due to a diffuse interface approximation. Hence, it might be
interesting to study the sharp interface limit in this context.
Moreover, Alberti and DeSimone connect nicely their generally valid homogenized formulas with the
classical results from \cite{Wenzel1936}
and \cite{Cassie1944}. In fact, they show that the Wenzel
and Cassie-Baxter laws represent upper bounds for the effective contact angle formula
derived in \cite{Alberti2005}.

There are of course open questions and future perspectives. A 
characterization of the effective macroscopic
Cahn-Hilliard equation by error estimates as exemplified in
different contexts in
\cite{Bensoussans1978,Schmuck2012} is of great
interest. Analytically, the
convergence of the microscopic (periodic) formulation to the
effective macroscopic Cahn-Hilliard problem is of great relevance.
In applications, it is very interesting to extend the porous media
formulation to fluid flow. It is well known that such an extension
is rather involved, since additional physical phenomena like
diffusion-dispersion effects arise (e.g. Taylor-Aris dispersion). It
is still not entirely clear how one can reliably account for such
phenomena.

Nevertheless, even without fluid flow, the new equations enable {\bf us} to
gain insight into interfacial dynamics in porous media for instance.
Two- or three-dimensional numerical results of wetting phenomena in
porous media would allow to track the phase interface of an
arbitrary three-phase composite, i.e. the porous medium and
arbitrary two phases in pore space. Such an information is of great
interest for the design of synthetic porous media, membranes, and
generally micro-fluidic devices. But the new formulation also
provides an interesting alternative for simulating oil recovery from
natural porous media.

\section*{Acknowledgements}
We thank the anonymous referees for insightful comments and
suggestions. We acknowledge financial support from EPSRC Grant No.
EP/H034587, EU-FP7 ITN Multiflow and ERC Advanced Grant No. 247031.

\bibliographystyle{rspublicnat}  
\bibliography{effWetting6} 
          %

\bigskip
\begin{center}
{\bf Appendix: Proof of Theorem \ref{thm:EfPhFi}}
\end{center}\label{sec:FoUp2}
We define the micro-scale $\frac{{\bf x}}{\epsilon}=:{\bf y}\in Y$ such that after setting,
\bsplitl{
\begin{array}{ll}
{\cg A}_0
	= 
	-\sum_{i,j=1}^d\pd{}{y_i}\brkts{\delta_{ij} \pd{}{y_j}}
	\,,
&\quad
{\cg B}_0
	= 
	- \sum_{i,j=1}^d\pd{}{y_i}\brkts{{\rm m}_{ij}\pd{}{y_j}}
	\,,
\\
{\cg A}_1
	= 
	-\sum_{i,j=1}^d\biggl[\pd{}{x_i}\brkts{\delta_{ij}\pd{}{y_j}}
&\quad
{\cg B}_1
	= 
	-\sum_{i,j=1}^d\biggl[\pd{}{x_i}\brkts{{\rm m}_{ij}\pd{}{y_j}}
\\\quad\quad
		+\pd{}{y_i}\brkts{\delta_{ij}\pd{}{x_j}}
		\biggr]\,,
&\qquad\quad
		+\pd{}{y_i}\brkts{{\rm m}_{ij}\pd{}{x_j}}
		\biggr]\,,
\\
{\cg A}_2
	= - \sum_{i,j=1}^d\pd{}{x_j}\brkts{\delta_{ij}\pd{}{x_j}}\,,
&\quad
{\cg B}_2
	= - \sum_{i,j=1}^d\pd{}{x_j}\brkts{{\rm m}_{ij}\pd{}{x_j}}\,,
\end{array}
}{B0B1B2}
${\cg A}_\epsilon := \epsilon^{-2}{\cg A}_0
+ \epsilon^{-1}{\cg A}_1 +{\cg A}_2$, and ${\cg B}_\epsilon:=\epsilon^{-2}{\cg B}_0
+ \epsilon^{-1}{\cg B}_1 +{\cg B}_2$, the Laplace operators $\Delta$ and ${\rm div}\brkts{\hat{\rm M}\nabla}$ 
become $\Delta u^\epsilon({\bf x}) =
{\cg A}_\epsilon u({\bf x}, {\bf y})$ and ${\rm div}\brkts{\hat{\rm M}\nabla}u^\epsilon({\bf x}) =
{\cg B}_\epsilon u({\bf x}, {\bf y})$, respectively, where $u^\epsilon({\bf x}):=u({\bf x}, {\bf y})$. Inserting for $u\in\brcs{w,\phi}$ the formal asymptotic expansions
$u^\epsilon
    \approx u_0({\bf x},{\bf y},t)
    +\epsilon u_1({\bf x},{\bf y},t)
    +\epsilon^2 u_2({\bf x},{\bf y},t)\,,$
%
into \reff{DePhMo0} and using \reff{B0B1B2} 
provides a sequence of three solvable perturbation problems, at ${\cal O}(\epsilon^{-2})$, 
${\cal O}(\epsilon^{-1})$ and ${\cal O}(\epsilon^{0})$, after equating terms of equal powers 
in $\epsilon$. For simplicity, we only give the last one here:
\bsplitl{
\mathcal{O}(\epsilon^{0}):\quad
\begin{cases}
{\cg B}_0w_2
    = -\lambda^2\brkts{
        {\cg B}_2 w_0
        +{\cg B}_1 w_1
    }
\\ \qquad\qquad
    -{\cg B}_0 \ebrkts{
            \frac{1}{2}f''(\phi_0)\phi_1^2
            +f'(\phi_0)\phi_2
        }
\\ \qquad\qquad
    -{\cg B}_1\ebrkts{
        f(\phi_0)\frac{\phi_1}{\phi_0}
    }
    -{\cg B}_2 f(\phi_0)
    -\partial_t(-\Delta)^{-1}w_0
    &\textrm{in }Y^1\,,
\\\quad
\textrm{no flux b.c.}\,,
\\\quad
\textrm{$w_2$ is $Y^1$-periodic}\,,
\\
{\cg A}_0 v_2
    = -{\cg A}_2 v_0
    -{\cg A}_1 v_1
    +w_0
    &\textrm{in }Y^1\,,
\\\quad
\nabla_n v_2
    = g_\epsilon
    &\textrm{on }\partial Y^1_w\,,
\\\quad
\textrm{$\phi_2$ is $Y^1$-periodic}\,,
\end{cases}
}{O-0n}
where in \reff{O-0n} the following relation is applied,
\bsplitl{
\frac{1}{2}f''(\phi_0)\phi_1^2
        +f'(\phi_0)\phi_2
        =
        a_1\phi_2
        +a_2\brkts{
            2\phi_2\phi_0
            +\phi_1^2
        }
        +3a_3\brkts{
            \phi_2\phi_0^2
            +\phi_0\phi_1^2
        }\,.
}{PoOrRe}

The first problems at ${\cal O}(\epsilon^{-2})$ are classical in elliptic homogenization theory 
and immediately imply that the
leading order approximations $w_0$ and $v_0$ are independent of the microscale ${\bf
y}$. This fact and the linear structure of the problems arising at ${\cal O}(\epsilon^{-1})$ suggest the
following ansatz for $w_1$ and $\phi_1$, i.e., 
\bsplitl{
w_1({\bf x},{\bf y},t)
    & = -\sum_{k=1}^d \xi^k_w({\bf y})\pd{w_0}{x_k}({\bf x},t)\,,
\quad
\phi_1({\bf x},{\bf y},t)
    = -\sum_{k=1}^d \xi^k_v({\bf y})\pd{\phi_0}{x_k}({\bf x},t)
    = v_1
    \,.
}{XiwXiphi}
Inserting \reff{XiwXiphi} into the ${\cal O}(\epsilon^{-1})$-problems provides equations for the correctors
$\xi^k_w$ and $\xi^k_v$. The resulting equation for $\xi^k_v$ is again
standard in elliptic homogenization theory and can be immediately written
for $1\leq k\leq d$ as,
\bsplitl{
\xi_\phi:\quad
\begin{cases}
-\sum_{i,j=1}^d
    \pd{}{y_i}\brkts{
        \delta_{ik}-\delta_{ij}\pd{\xi^k_v}{y_j}
    }
    =
    &
\\\qquad\quad\,\,
    = -{\rm div}\brkts{
        {\bf e}_k-\nabla_y\xi^k_v
    }=0
    &\textrm{ in }Y^1\,,
\\
        {\bf n}\cdot\brkts{
            \nabla\xi^k_v
            +{\bf e}_k
        }
    =
    0
    &\textrm{ on }\partial Y^1_w\,, 
\\
\xi^k_v({\bf y})\textrm{ is $Y$-periodic and ${\cg M}_{Y^1}(\xi^k_v)=0$.}
\end{cases}
}{Xiphi}

The reference cell problem for $\xi_w$ is much more difficult since it
depends on the solutions of \reff{Xiphi}. We first write the problem at ${\cal O}(\epsilon^{-1})$ 
for $\xi^k_w$ in explicit terms,
\bsplitl{
& \sum_{k,i,j=1}^d
    \pd{}{y_i}\brkts{
        {\rm m}_{ij}\pd{\xi^k_w}{y_j}
    }\pd{w_0}{x_k}
    =
    \sum_{i,j=1}^d\pd{}{y_i}\brkts{
        {\rm m}_{ij}\pd{w_0}{x_j}
    }
\\&\qquad\quad
    -\frac{f(\phi_0)}{\phi_0}\sum_{k,i,j=1}^d\pd{}{y_i}\brkts{
        {\rm m}_{ij}\pd{\xi_v^k}{y_j}
    }\pd{\phi_0}{x_k}
    +\sum_{k,i,j=1}^d\pd{}{y_i}\brkts{
        f'(\phi_0){\rm m}_{ij}\pd{\phi_0}{x_i}
    }\,.
}{XiwA} At this point a major problem is the dependence on $\phi_0$
in problem \reff{Xiw}. To alleviate this difficulty, we make use of 
the chemical potential defined in \reff{LoEqCo}.
In the case of thermodynamic equilibrium, the quantity $\mu$
is constant. Hence, it holds that,
$f'(\phi)\pd{\phi}{x_k}
    = f'(\phi)\pd{v}{x_k}
    =\lambda^2\pd{w}{x_k}\
    \quad\textrm{for }1\leq k\leq d\,.
$ 
If this identity is valid in each reference cell $Y$ (that means, locally)
and the mobility tensor $\hat{\rm M}$ is isotropic, i.e.
$\hat{\rm M}=\brcs{{\rm m}_{ij}}_{1\leq i,j\leq d}=\brcs{{\rm m}\delta_{ij}}_{1\leq i,j\leq d}$, then
we can cancel $\pd{w_0}{x_k}$ in \reff{XiwA} and simplify to,
\bsplitl{
\begin{cases}
-\sum_{i,j,k=1}^d
    \pd{}{y_i}\brkts{
        \delta_{ik}-\delta_{ij}\pd{\xi^k_w}{y_j}
    }
\\\qquad\qquad
    =
    \lambda^2 \sum_{k,i,j=1}^d\pd{}{y_i}\brkts{
        {\rm m}_{ik}
        -\frac{f(\phi_0)}{f'(\phi_0)\phi_0}{\rm m}_{ij}\pd{\xi^k_v}{y_j}
    }
    &\textrm{ in }Y^1\,,
\\
\sum_{i,j,k=1}^d{\rm n}_i\Bigl(
        \brkts{
        \delta_{ij}\pd{\xi^k_w}{y_j}
        -\delta_{ik}
        }
\\\qquad\qquad
        -\lambda^2 \sum_{k,i,j=1}^d\pd{}{y_i}\brkts{
            {\rm m}_{ik}
            -\frac{f(\phi_0)}{f'(\phi_0)\phi_0}{\rm m}_{ij}\pd{\xi^k_v}{y_j}
        }
    \Bigr)
    = 0
    &\textrm{ on }\partial Y^1_w\,, 
\\
\xi^k_w({\bf y})\textrm{ is $Y$-periodic and ${\cg M}_{Y^1}(\xi^k_w)=0$.}
\end{cases}
}{Xiw}
One can guarantee well-posedness of the cell problem \reff{Xiw} under the Assumption {\bf F} 
which ensures that $r(s):=f(s)/(f'(s)s)\in L^2([\alpha_1,\alpha_2])$ and since $\phi_0\in H^2_E(\Omega)$, it holds that
\bsplitl{
\int_\Omega r^2(\phi_0)\,d{\bf x}
	\leq \av{\Omega}\int_{\alpha_1}^{\alpha_2}r^2(s)\,ds
	<\infty\,,
}{Wllpsd}
such that $\xi^k_w\in L^2(\Omega;H^1_{per}(Y^1))$.

We come to the last problem \reff{O-0n}. Again, equation
\reff{O-0n}$_2$ is much simpler because it is standard in elliptic
homogenization theory. Well-known existence and uniqueness results
(Fredholm alternative/Lax-Milgram) immediately guarantee solvability
by verifying that the right hand side in \reff{O-0n} is zero as an
integral over $Y^1$. 
For $\tilde{g}_0:=-\frac{\gamma}{C_h}\int_{\partial Y^1}\brkts{a_1\chi_{\partial Y^1_{w_1}}
    +a_1\chi_{\partial Y^1_{w_2}}}\,do({\bf y})$
we obtain the following
effective equation for the phase field,
\bsplitl{
-\sum_{i,k=1}^d\ebrkts{
        \sum_{j=1}^d\int_{Y^1}\brkts{
            \delta_{ik}-\delta_{ij}\pd{\xi^k_v}{y_j}
        }\,d{\bf y}
    }\pd{^2v_0}{x_i\partial x_k}
    = \av{Y^1}w_0
    +\tilde{g}_0
    \,,
}{UpPhi1}
which can be written more compactly by defining a porous media correction tensor
$\hat{\rm D}:=\brcs{{\rm d}_{ik}}_{1\leq i,k\leq d}$ by
\bsplitl{
\av{Y}{\rm d}_{ik}
    := \sum_{j=1}^d\int_{Y^1}\brkts{
        \delta_{ik}
        -\delta_{ij}\pd{\xi^k_v}{y_j}
    }\,d{\bf y}\,.
}{Dphi}
Equations \reff{UpPhi1} and \reff{Dphi} provide the final form of the upscaled
equation for $\phi_0$, i.e.,
$-\Delta_{\hat{\rm D}} v_0
    :=
    -{\rm div}\brkts{
        \hat{\rm D}\nabla v_0
    }
    = \theta_1w_0
    +\tilde{g}_0\,.$

The upscaled equation for $w$ is again a result of the Fredholm alternative, i.e., a solvability
criterion on equation \reff{O-0n}$_1$. 
This means that we require,
\bsplitl{
\int_{Y^1}\Bigl\{
    -\lambda^2\brkts{
         {\cg B}_2 w_0
        +{\cg B}_1 w_1
    }
    -{\cg B}_0 \brkts{
        \frac{1}{2}f''(\phi_0)\phi_1^2
        +f'(\phi_0)\phi_2
    }
\\ \qquad\qquad
    -{\cg B}_1\ebrkts{
        f(\phi_0)\frac{\phi_1}{\phi_0}
    }
    -{\cg B}_2f(\phi_0)
    -\partial_t(-\Delta)^{-1}w_0
    \Bigr\}\,d{\bf y}
    =0\,.
}{SoCow2}
Let us start with the terms that are easily averaged over the reference
cell $Y$. The first two terms in \reff{SoCow2} can be rewritten by,
\bsplitl{
\int_{Y^1}-\brkts{
        {\cg B}_2 w_0
        +{\cg B}_1 w_1
    }\,d{\bf y}
    =
    -\sum_{i,k=1}^d\ebrkts{
        \sum_{j=1}^d\int_{Y^1}\brkts{
            {\rm m}_{ik}-{\rm m}_{ij}\pd{\xi^k_w}{y_j}
        }\,d{\bf y}
    }\pd{^2w_0}{x_i\partial x_k}
\\
    = -{\rm div}\brkts{
        \hat{\rm M}_w\nabla w_0
    }
    \,,
}{Bw}
where the effective tensor $\hat{\rm M}_w=\brcs{{\rm m}^w_{ik}}_{1\leq i,k\leq d}$ is defined by
\bsplitl{
{\rm m}^w_{ik}
    & :=
    \frac{1}{\av{Y}}\sum_{j=1}^d\int_{Y^1}\brkts{
        {\rm m}_{ik}
        -{\rm m}_{ij}\pd{\xi^k_w}{y_j}
    }\,d{\bf y}\,.
}{Mw}
The next terms in \reff{SoCow2} become
\bsplitl{
-{\cg B}_1\ebrkts{
        f(\phi_0)\frac{\phi_1}{\phi_0}
    }
    & -{\cg B}_2f(\phi_0)
    =
    \sum^d_{k,i,j=1}\biggl\{
        -\pd{}{x_i}\brkts{
            \ebrkts{
                {\rm m}_{ij}\frac{f(\phi_0)}{\phi_0}\pd{\xi^k_v}{y_j}
            }\pd{\phi_0}{x_k}
        }
\\&
        +\pd{}{y_i}\brkts{
            {\rm m}_{ij}\frac{f(\phi_0)}{\phi_0}\pd{\phi_1}{x_j}
        }
        +\pd{}{y_i}\brkts{
            {\rm m}_{ij}\phi_1\pd{(f(\phi_0)/\phi_0)}{x_j}
        }
    \biggr\}
\\&
    +\sum_{k,i,j=1}^d
        \pd{}{x_i}\brkts{
            {\rm m}_{ij}f'(\phi_0)\pd{\phi_0}{x_j}
        }
    \,,
}{Bphi}
and a subsequent integration of the right hand side of \reff{Bphi}
over the reference cell $Y$ gives
\bsplitl{
    & \sum^d_{i,k=1}\pd{}{x_i}
    \brkts{
        \ebrkts{
            \sum_{j=1}^d \int_{Y^1}\brkts{
                {\rm m}_{ik}f'(\phi_0)-{\rm m}_{ij}\frac{f(\phi_0)}{\phi_0}\pd{\xi^k_v}{y_j}
            }\,d{\bf y}
        }\frac{\partial \phi_0}{\partial x_j}
    }
\\&
    -\sum_{k,j=1}^d\ebrkts{
        \sum_{i=1}^d
        \int_{Y^1}\brkts{
            {\rm m}_{ij}\frac{\partial \xi^k_v}{\partial y_i}
        }\,d{\bf y}
    }\frac{f(\phi_0)}{\phi_0}\frac{\partial^2\phi_0}{\partial x_k\partial x_j}
\\&
    -\sum_{k,j=1}^d\ebrkts{
        \sum_{i=1}^d\int_{Y^1}\brkts{
            \pd{\xi^k_v}{y_i}{\rm m}_{ij}
        }\,d{\bf y}
    }
    \frac{\partial (f(\phi_0/\phi_0)}{\partial x_j}\frac{\partial \phi_0}{\partial x_k}
    \,,
}{Bphi2}
where the last two terms further simplify to
\bsplitl{
    -\sum_{k,j=1}^d
    \pd{}{x_j}\brkts{
        \frac{f(\phi_0)}{\phi_0}
            \ebrkts{\sum_{i=1}^d
            \int_{Y^1}\brkts{
                {\rm m}_{ij}\frac{\partial \xi^k_v}{\partial y_i}
            }\,d{\bf y}
            }
        \frac{\partial \phi_0}{\partial x_k}
    }
    \,.
}{Bphi3}
With \reff{Bphi3} we can finally write \reff{Bphi} in the following compact way
\bsplitl{
\frac{1}{\av{Y}}\int_{Y^1}\brkts{
    -{\cg B}_1\ebrkts{
            f(\phi_0)\frac{\phi_1}{\phi_0}
        }
         -{\cg B}_2f(\phi_0)
    }\,d{\bf y}
\\
    = \sum^d_{i,k=1}\pd{}{x_i}\brkts{
        \ebrkts{
            \frac{1}{\av{Y}}\sum_{j=1}^d \int_{Y^1}\brkts{
                {\rm m}_{ik}f'(\phi_0)-2{\rm m}_{ij}\frac{f(\phi_0)}{\phi_0}\pd{\xi^k_v}{y_j}
            }\,d{\bf y}
        }\frac{\partial \phi_0}{\partial x_j}
    }
\\
    =
    {\rm div}\brkts{
        \ebrkts{
            \theta_1f'(\phi_0)\hat{\rm M}
            -2\frac{f(\phi_0)}{\phi_0}\hat{\rm M}_v
        }\nabla \phi_0
    }
    \,,
}{Bphi4}
where the tensor $\hat{\rm M}_v=\brcs{{\rm m}^v_{ij}}_{1\leq i,k\leq d}$ is defined by
\bsplitl{
{\rm m}^v_{ik}
    & :=
    \frac{1}{\av{Y}}\sum_{j=1}^d\int_{Y^1}\brkts{
        {\rm m}_{ij}\pd{\xi^k_v}{y_j}
    }\,d{\bf y}\,.
}{Mv}
It remains to elucidate the last term in \reff{SoCow2}. Using \reff{O-0n}$_2$,
then we have,
\bsplitl{
    -{\cg B}_0 \ebrkts{
        \frac{1}{2}f''(\phi_0)\phi_1^2
        +f'(\phi_0)\phi_2
    }
    =
    \sum_{i,j=1}^d\pd{}{y_i}\brkts{
        {\rm m}_{ij}\phi_1\pd{\phi_1}{y_j}
    }f''(\phi_0)
\\
    +\sum_{i,j=1}^d\pd{}{y_i}\brkts{
        {\rm m}_{ij}\pd{\phi_2}{y_j}
    }f'(\phi_0)
\\
    =
    \sum_{i,j=1}^d-\ebrkts{
        \pd{}{y_i}\brkts{
            {\rm m}_{ij}\phi_1\pd{\xi^k_v}{y_j}
        }
    }f''(\phi_0)\pd{\phi_0}{x_k}
    +{\rm m}\brkts{
        {\cg A}_2\phi_0
        +{\cg A}_1\phi_1
        -w_0
    }\,.
}{B0Phi}
If we assume an isotropic mobility matrix
$\hat{\rm M}$, i.e., $\hat{\rm M}=\brcs{{\rm m}\delta_{ij}}_{1\leq i,j\leq d}$, and
use \reff{Xiphi} in the term with the summation, then the following
simplification of its summands can be made,
\bsplitl{
\ebrkts{\pd{}{y_i}\brkts{
            {\rm m}_{ij}\phi_1\pd{\xi^k_v}{y_j}
        }
    }f''(\phi_0)\pd{\phi_0}{x_k}
    =
    \ebrkts{
        \pd{}{y_i}\brkts{
            {\rm m}_{ik}\phi_1
        }
    }\pd{f'(\phi_0)}{x_k}
    =
    -\ebrkts{
        {\rm m}_{ik}\pd{\xi^l_v}{y_i}
    }\pd{\phi_0}{x_l}\pd{f'(\phi_0)}{x_k}
\\
    =
    -\pd{}{x_k}\brkts{
        f'(\phi_0)\ebrkts{
            {\rm m}_{ik}\pd{\xi_v^l}{y_i}
        }\pd{\phi_0}{x_l}
    }
    +f'(\phi_0)\pd{}{x_k}\brkts{
        \ebrkts{
            {\rm m}_{ik}\pd{\xi_v^l}{y_i}
        }\pd{\phi_0}{x_l}
    }
    \,.
}{Simp}
The last term in \reff{B0Phi} vanishes by the Fredholm alternative guaranteeing
solvability of \reff{O-0n}$_2$. 
Hence, \reff{B0Phi} admits after integrating over $Y$ the
following compact form,
\bsplitl{
\frac{1}{\av{Y}}\int_{Y^1}
        -{\cg B}_0 \ebrkts{
        \frac{1}{2}f''(\phi_0)\phi_1^2
        +f'(\phi_0)\phi_2
    }\,d{\bf y}
    =
    {\rm div}\brkts{
        f'(\phi_0)\hat{\rm M}_v\nabla \phi_0
    }\quad
\\
    -f'(\phi_0){\rm div}\brkts{
        \hat{\rm M}_v\nabla \phi_0
    }
    \,,
}{B0Phi2}
which then sets \reff{B0Phi} to zero. These considerations finally lead to
the following effective equation for $\phi_0$, i.e.,
\bsplitl{
\theta_1\pd{\phi_0}{t}
    = {\rm div}\brkts{
        \ebrkts{
            \theta_1f'(\phi_0)\hat{\rm M}
            -\brkts{2\frac{f(\phi_0)}{\phi_0} 
            - f'(\phi_0)
            }\hat{\rm M}_v
        }\nabla \phi_0
    }
\\
    -f'(\phi_0){\rm div}\brkts{
        \hat{\rm M}_v\nabla \phi_0
    }
    +\frac{\lambda^2}{\theta_1}{\rm div}\brkts{
        \hat{\rm M}_w\nabla \brkts{
            {\rm div}\brkts{
                \hat{\rm D}\nabla \phi_0
            }
            -\tilde{g}_0
        }
    }\,.
}{EfW0}
The solvability of \reff{EfW0} follows along with the arguments in \cite{Novick-Cohen1990} since we at least assume
that $f\in C^2_{Lip}(I)$ where $I\subset\mathbb{R}$ is a bounded interval. In fact, one only needs to prove a local Lipschitz continuity of the first twoterms on the right hand side of \reff{EfW0}.

\end{document}